\documentclass[pra,twocolumn,showpacs,preprintnumbers,superscriptaddress,amsmath,amssymb]{revtex4}

\usepackage{graphicx}
\usepackage{dcolumn}
\usepackage{bm}

\usepackage{amsmath}	
\begin{document}
\draft
\newcommand{\lw}[1]{\smash{\lower2.ex\hbox{#1}}}

\title{Correlation Effects on Atom Density Profiles of 1-D and 2-D Polarized Atomic-Fermi-Gas Loaded on Optical Lattice}

\author{M.~Machida} 
\email{machida.masahiko@jaea.go.jp}
\affiliation{CCSE, Japan Atomic Energy Agency, 6-9-3 Higashi-Ueno,
Taito-ku Tokyo 110-0015, Japan}
\affiliation{CREST(JST), 4-1-8 Honcho, Kawaguchi, Saitama 332-0012,
Japan}
\author{S.~Yamada}
\email{yamada.susumu@jaea.go.jp}
\affiliation{CCSE, Japan Atomic Energy Agency, 6-9-3 Higashi-Ueno,
Taito-ku Tokyo 110-0015, Japan}
\affiliation{CREST(JST), 4-1-8 Honcho, Kawaguchi, Saitama 332-0012,
Japan}
\author{M.~Okumura}
\email{okumura.masahiko@jaea.go.jp}
\affiliation{CCSE, Japan Atomic Energy Agency, 6-9-3 Higashi-Ueno,
Taito-ku Tokyo 110-0015, Japan}
\affiliation{CREST(JST), 4-1-8 Honcho, Kawaguchi, Saitama 332-0012,
Japan}
\author{Y.~Ohashi}
\email{yohashi@rk.phys.keio.ac.jp}
\affiliation{Faculty of Science and Technology, Keio University, 3-14-1,
Hiyoshi, Kohoku-ku, Yokohama, Kanagawa, 223-0061, Japan}
\affiliation{CREST(JST), 4-1-8 Honcho, Kawaguchi, Saitama 332-0012,
Japan}
\author{H.~Matsumoto}
\email{mhideki@mx1.ttcn.ne.jp}
\affiliation{IMR, Tohoku University, 2-1-1, Katahira, Aoba-ku, Sendai,
Miyagi 980-8577, Japan}
\affiliation{CREST(JST), 4-1-8 Honcho, Kawaguchi, Saitama 332-0012,
Japan}

\date{\today}

\begin{abstract} 

 We investigate effects of optical lattice potential in one- and
 two-dimensional two-component trapped Fermi gases with population
 imbalances. Using the exact diagonalization and the density matrix
 renormalization group methods complementarily, we calculate the atom
 density profile from the ground state many-body wavefunction as a
 function of attractive interaction strength for various population
 imbalances. The numerical results reveal that although a phase
 separation between the superfluid core and the shell cloud of excess
 atoms occurs as observed in experiments without the optical lattice,
 the population imbalance generally remains in the core region in
 contrast to the non-lattice cases. The essence of the numerical results
 in a strong attractive regime can be explained by an effective model
 composed of Cooper pairs and excess major fermions.  

\end{abstract}


\maketitle

\section{Introduction}
%
%
Recently, two-component fermion systems with population imbalance have 
attracted much attention in various research fields, such as cold atoms,
superconductors, and QCD \cite{REVM}. In the 1960's, effects of the
population imbalance have been theoretically investigated in the
superconductivity literature \cite{Sarma,FF,LO,Liu}. Sarma considered
the stability of the gap phase (Sarma state) \cite{Sarma} and Liu and
Wilczek revisited it with a new picture (interior gap phase)
\cite{Liu}. Fulde and Ferrell \cite{FF}, and Larkin and Ovchinikov
\cite{LO} predicted the so-called FFLO state, where the superconducting
order parameter is spatially modulated. Very recently, some evidence of
the FFLO state has been reported in a heavy fermion superconductor ${\rm
CeCoIn}_5$ \cite{Kumagai}. 

The population imbalance has been also extensively studied in ultra-cold
Fermi gases \cite{Zwie1,Zwie2,Shin1,Part1,Part2,Schunck}. The advantage
of using atom gases is that one can widely tune some physical
parameters, such as the interaction and the population imbalance. In a
trapped two-component $^6{\rm Li}$ Fermi gas
\cite{Zwie1,Zwie2,Shin1,Part1,Part2}, a phase separation between a
superfluid core region and a surrounding unpaired gas of excess atoms
was observed. Recently, the absence of the population imbalance inside
the superfluid core region was reported in \cite{Shin1,Part2}.   

So far, the superfluid Fermi gas with population imbalance has been
experimentally studied without an optical lattice
\cite{Zwie1,Zwie2,Shin1,Part1,Part2,Schunck}. In this paper, we
investigate effects of the lattice
\cite{Greiner,Kohl,SuperOptLatt,OptLattReview} on the population
imbalanced systems. Very recently, there have been several reports on
this issue, one of which theoretically claims that FFLO is more
stabilized than the non-optical lattice case \cite{Torma}, and others
of which numerically show that FFLO is observable in the one-dimensional
attractive Hubbard model \cite{DMRGFF} by using the density matrix
renormalization Group (DMRG) method \cite{White,DMRGReview}. Our
approach in this paper differs from these works in using the exact
diagonalization method mainly and the DMRG method
\cite{White,DMRGReview} complementarily in order to check whether the
exact diagonalization results are small size effects or not. In
addition, we focus on the density profile of majority ($n_\uparrow$) and
minority components ($n_\downarrow$) at $T=0$ in a one- and
two-dimensional lattice trapped systems instead of the stability of the
exotic superfluidity FFLO. The reason is that the atom density profile
is the most convenient and clear observable for experiments. Moreover,
the FFLO requires at least the population imbalance in the atom density
profile. Namely, it is crucial point for FFLO whether the population
imbalance survives even in the presence of pair instability at $T=0$
ground state or not. Thus, we widely measure the atom density profiles
in various system sizes, imbalance, interaction, dimensionality (1D-2D),
and traps, prior to an examination of the FFLO stability. Consequently,
we point out in a strongly-attractive interaction regime that
correlation effects due to the presence of lattice result in a core
phase with a typical population imbalance, which is very different from
non-lattice cases \cite{Shin1,Part2}. 

The contents of this paper are as follows. In Section \ref{PDP}, we
summarize the experimental results of the atom density profiles in the
imbalanced Fermi gases without the optical lattice and show our
prediction on those of the imbalanced gases in the presence of the
optical lattice. Section \ref{MNCM} is devoted to explain the model and
the numerical method. In Section \ref{NR1D}, numerical results by the
exact diagonalization and DMRG are presented, theoretical analysis using
an effective model derived in strong coupling limit is given, and strong
trap effects are displayed as an exceptional case. Section \ref{NR2D}
shows the results in two dimensional (2-D) systems, and Section
\ref{TempE} discusses temperature effects on all the present
results. Finally, summary and conclusion are given. 

\section{Particle Density Profiles in Polarized Atomic Fermi
 Gases\label{PDP}}

\begin{figure}
\includegraphics[scale=0.4]{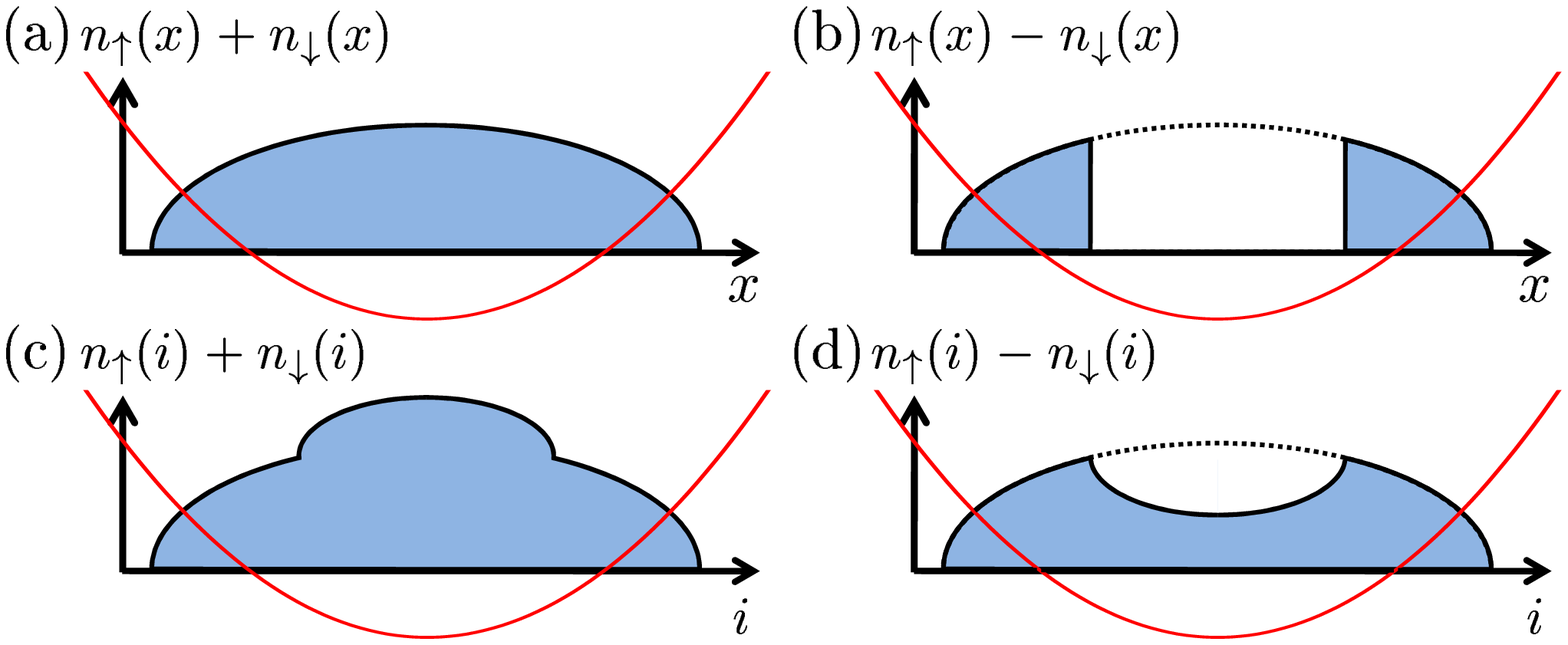}
\caption{\label{fig1} Schematic profiles of the total atom density and 
the density difference between particles with pseudo-spin up and down,
 i.e., $n_\uparrow - n_\downarrow $, which are measured along the long
 axis of the trap for polarized fermionic gases. The upper panels (a)
 and (b) display the experimental (non-lattice system) results, while
 the lower panels (c) and (d) predict the profiles in the presence of
 the optical lattice. The parabolic lines schematically represent the
 trap potential shape. In the absence of the optical lattice, the
 experiments revealed that the phase separation between the paired
 (superfluid) gas and the completely polarized (excess major atoms only)
 periphery occurs as shown in the upper panels (a) and (b). On the other
 hand, the lower panels (c) and (d) show a typical result of numerical
 calculations in the presence of the optical lattice. A big difference
 between the upper and the lower panels is that the excess species can
 stay even on the superfluid core in the presence of the optical
 lattice.} 
\end{figure}

In this section, we summarize the experimental results
\cite{Shin1,Part2} of the trapped imbalanced Fermi gases without the
optical lattice and predict how their results change by adding the
optical lattice. Figure \ref{fig1} (a) and (b) are schematic pictures of
atomic density profiles obtained in experiments without the optical
lattice, and (c) and (d) are our prediction of those in the presence of
the optical lattice. At first, let us focus on Fig.~\ref{fig1} (a) and
(b). The experiments revealed that the polarized gases show a phase
separation between the core and the periphery below the superfluid
transition temperature. The phase separation picture was clearly
confirmed by monitoring the atom density profile for each of two
Fermi-atom species. Inside the core, the profile shows an atom density
ratio, $n_\uparrow (x) :n_\downarrow (x) =1:1$ as shown in
Fig.~\ref{fig1} (b) \cite{Shin1,Part2}, while a completely-polarized
phase appears in the periphery, where the atom density ratio,
$n_\uparrow (x) :n_\downarrow (x) =1:0$ in a case that the majority
species is Fermi atom with the pseudo-spin $\uparrow$.
 
On the other hand, our calculations in the presence of the optical
lattice reveal that although the phase separation occurs like the
experimental results without the optical lattice, the atom density
profile shows a big difference compared to the non-lattice cases. We
find in the presence of lattice that the population imbalance generally
remains even in the superfluid core region, i.e., the atom density ratio
inside the core, $n_\uparrow (i) :n_\downarrow (i) \ne 1:1$. Moreover,
we find in sufficiently strong pairing-interaction that $n_\uparrow (i)
:n_\downarrow (i) \simeq 2:1$ \cite{NoteD} inside the core as shown in
Fig.~\ref{fig1} (d). This result is compared with an effective model
theoretically derived in the strong limit of the attractive
interaction. In this paper, we claim that correlation effects on the
lattice bring about the imbalanced core phase. The main correlation
effects on the lattice originate from a repulsive interaction between
two Cooper pairs \cite{Machida1} and a site-exchange energy gain between
a Cooper pair and an excess fermion as shown in \ref{NR1D}. The
effective model successfully explains why the imbalanced core emerges. 
\par


\section{Model and Numerical Calculation Method\label{MNCM}}


Let us show the model Hamiltonian and the numerical calculation method
to obtain the ground state of the model. In this paper, we consider
two-component trapped Fermi gases inside optical lattices with $N$
lattice sites. When the optical lattice potential is sufficiently
strong, the system can be well described by the Hubbard model, the
one-dimensional (1-D) Hamiltonian of which is given as
\cite{Rigol,Machida}, 
\begin{align}
H_{\rm Hubbard} & = - t \sum_{\langle i,j \rangle, \sigma} ( a_{j
\sigma}^\dag a_{i \sigma} + {\rm H.c.}) + U \sum_i  n_{i \uparrow} n_{i
\downarrow} \nonumber \\  
& \quad {} + V {\left({\frac{2}{N-1}}\right)}^2 \sum_{i,\sigma} n_{i
\sigma} \left(i-\frac{N+1}{2}\right)^2 , \label{eq.1}
\end{align}
where $a_{i\sigma}^\dag$ is the creation operator of a Fermi particle
with pseudo-spin $\sigma=\uparrow$ or $\downarrow$ at $i$-th site, and
$U$ ($<0$) is a pairing interaction. The summation $\langle i,j \rangle$
is taken over the nearest-neighbor sites, and the hopping matrix element
$t$ describes the hopping energy of atoms between the nearest-neighbor
sites. The last term in Eq.~(\ref{eq.1}) gives a harmonic trap
potential, in which $V$ corresponds to the potential height at the edge
of the lattice. At the edge, the open-boundary condition is imposed. If
the atomic density profile drops down at the edge, then the boundary
condition does not almost affect the result. The parameter $V$ and the
total number of Fermi atoms $N_{\rm F}$ are selected to sufficiently
reduce the atom density at the edges.  

In this paper, we examine various population imbalances with keeping
$N_{\uparrow}  \geq N_{\downarrow}$ ($N_{\rm F} = N_{\uparrow} +
N_{\downarrow}$), where $N_{\uparrow}$ and $N_{\downarrow}$ are the
total number of Fermi atoms with pseudo-spin $\uparrow$ and
$\downarrow$, respectively. In 1-D cases, we numerically diagonalize the
Hamiltonian, Eq.~(\ref{eq.1}) to calculate the density profile at $T=0$. 
Although this approach gives us exact results, the accessible system
size is severely limited. To compensate this disadvantage and confirm
whether the exact diagonalization results are small size effects or not,
we employ the DMRG method. The DMRG guarantees a high accurate result as
long as the trap potential is not so steep. We confirm that a difference
between the eigen-values obtained by both methods is below $10^{-6}$ in
the lattice size $N=20$ with $V/t=1$. Since the DMRG is mainly used for
much less steep traps ($V/t$ is fixed to be 1 except for Section VI D,
which examines a trap strength dependence), the DMRG results
sufficiently keep high accuracy. 

On the other hand, in two-dimensional (2-D) square lattice systems, we
use the exact diagonalization solely, because the method is now the most
accurate and reliable for 2-D finite systems. Here, we just mention that
the direct extension of the DMRG to 2-D ladder cases, which keeps high
accuracy like 1-D cases is now under development \cite{YOM,MOY}. For the
exact diagonalization and the DMRG, we use SX-6 (1-node 4 CPU's system)
and Altix 3700Bx2 (scalar parallel machine) in JAEA and the Earth
Simulator \cite{Yamada}. In the exact diagonalization, the problems with
the lattice size $N=16$ are calculated by SX-6, while those with above
$N=20$ the Earth Simulator. In the latter case, we need parallelization
and high-performance computing techniques. See Ref.~\cite{Yamada} for
computational technical issues on massively parallel supercomputers. On
the other hand, all DMRG calculations are made on Altix 3700Bx2.  
  

\section{Numerical Results for 1-D Imbalanced Gases\label{NR1D}}

\subsection{Exact Diagonalization Studies}

\begin{figure}
\includegraphics[scale=0.34]{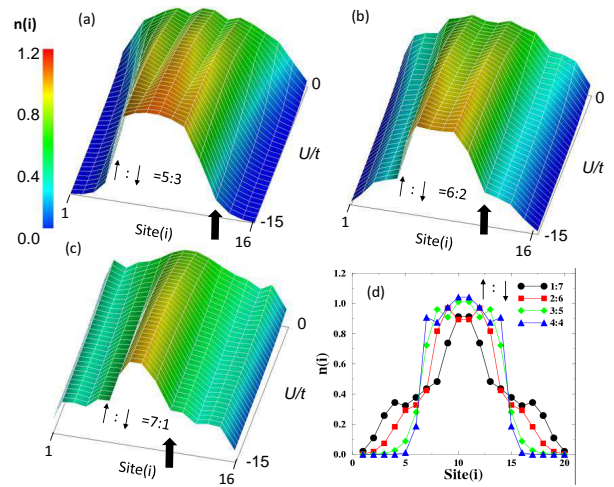}
\caption{\label{fig2} The exact diagonalization results for $U/t$
 dependences ($-15 \leq U/t \leq 0$) of particle density profiles,
 $n(i)$ for three imbalance cases, whose ratios of major to minor
 fermion species are (a) 5:3, (b) 6:2, and (c) 7:1, respectively. In
 these three cases, $N_{\rm F}=8$, $N=16$ and $V/t = 1$. The panel (d)
 is a highlight of $n(i)$ at $U/t=-10$ for the imbalance from 4:4 to 7:1
 in $N=20$. $N_{\rm F}$ and $V/t$ are the same as (a-c).} 
\end{figure}

Let us present exact diagonalization results for the imbalanced 1-D
attractive Hubbard model with the trap potential. Figure \ref{fig2}
shows $U/t$ dependences of the calculated atom-density profile of the
ground state, $n(i)=n_\uparrow(i)+n_\downarrow(i)$ for three population
imbalances, whose ratios of major to minor species are (a) 5:3, (b) 6:2,
and (c) 7:1, respectively. For these three cases, the lattice size
$N=16$, the total number of fermions $N_{\rm F} =8$, and $V/t=1$. The
filling is just the quarter one if the trap potential is
absent. Although the depression at the edges is not completely zero in
the presence of the trap ($V/t=1$), we confirm that there are no edge
effects because of no significant change of profiles on the lattice size
extension (to $N=18$ and $20$) with the same fermion number (see
Fig.~\ref{fig2} (d)).   

In each panel, we find a step-like structure (see arrow in each figure)
in each dome like profile. This structure becomes more pronounced as one
increases the amplitude strength of the attractive interaction
$|U|$. The structure is also observable in Fig.2(d) with a larger size
($N=20$), in which $N_{\rm F}$ and $V/t$ is the same as (a-c), and $U/t$
is fixed to be $-10$.  

\begin{figure}
\includegraphics[scale=0.33]{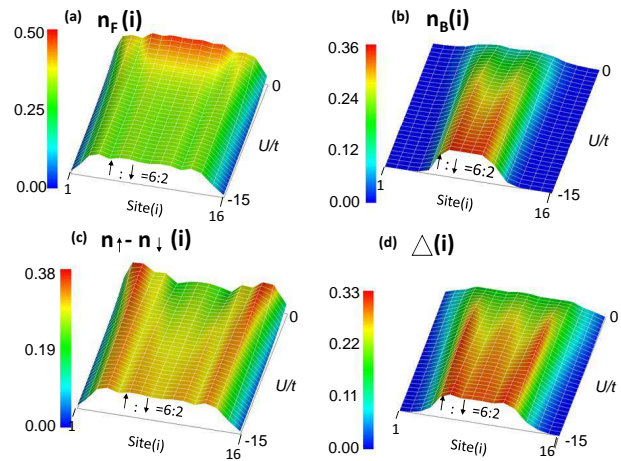}
\caption{\label{fig3} The exact diagonalization results for $U/t$
dependences ($-15 \leq U/t \leq 0$) of profiles of (a) the single
occupation density $n_{\rm F}(i)$, (b) the double occupation density
$n_{\rm B}(i)$, (c) the subtraction from major to minor species
density $n_{\uparrow} (i) - n_{\downarrow} (i)$, and (d) the on-site
pair amplitude $\Delta(i)$. In all the cases, $N_{\rm F} = 8$ ($ 6
\uparrow, 2 \downarrow$), and other conditions are the same as those of
Fig.~\ref{fig1} (a-c).} 
\end{figure}

The step-like structures seen in Fig.~\ref{fig2} are found to be
associated with the phase separation between a core phase whose main
component is pair and a shell one composed of only un-paired excess
atoms from other observables as shown in Fig.~\ref{fig3}. Figure
\ref{fig3} (a), (b), (c) and (d), respectively, show $U/t$ dependent
profiles of the single-occupation density $n_{\rm F}(i)$, the
double-occupation density $n_{\rm B}(i)$, the density subtraction
$n_{\uparrow}(i) - n_{\downarrow} (i)$, and the on-site pair amplitude
$\Delta(i)$ \cite{Machida,PairA} defined by  
\begin{align}
\Delta(i) & = \frac{1}{\sqrt{2}} \langle N_\uparrow+1,N_\downarrow + 1 |
a_{i \uparrow}^\dag a_{i \downarrow}^\dag | N_\uparrow,N_\downarrow
\rangle \, \label{eq.2}
\end{align}
where, $| N_\uparrow, N_\downarrow \rangle$ is the ground state
wavefunction with $(N_\uparrow, N_\downarrow)$ atoms. This function is a
probe for the local condensate amplitude \cite{PairA} for finite
systems. Here, we note that the probe gives non-zero values even at the
interaction free $(U/t =0)$ as seen in \ref{fig3} (d) and shows spatial
modulations without sign changes especially in a relatively-weak
interaction regime. The reason for the former one is that the function
counts all the possible pair formation probabilities on a local site,
i.e., it includes even the pair formation due to the trap potential, and
that for the latter one may be that the strong confinement effect masks
the sign change due to the FFLO. Although this latter interpretation
requires further examinations, we confirm at least the growth of the
local condensate amplitude. See Ref.\cite{DMRGFF} for the spatial pair
correlation in larger core phases. $n_{\rm F}(i)$ and $n_{\rm B}(i)$ are
calculated by picking up the coefficient of the eigen-state whose $i$-th
site occupation is 1 and 2, respectively, from the ground state
wave-function and summing up the amplitude of the coefficients. Their
explicit operator definitions are given by $n_{F} (i)  = < n_{\uparrow}
(i) + n_{\downarrow} (i) > -  n_{B} (i) $, where $n_B (i) = <
n_{\uparrow} (i) n_{\downarrow} (i)  >$. For the methodological details
to calculate $n_{\rm F}(i)$ and $n_{\rm B}(i)$, see
Ref.~\cite{Machida1}. According to Ref.~\cite{Machida1}, as $|U/t|$
increases in the balanced case, $n_{\rm F}(i)$ decreases, while $n_{\rm
B}(i)$ instead increases. This simple relationship means that the Cooper
pair becomes more tightly-bounded one on a site with increasing $|U/t|$. 

From Fig.~\ref{fig3}, we clearly find that $n_{\rm B}(i)$ and
$\Delta(i)$ well develop only inside the ``phase boundary''
characterized by the step-like structure as seen in Fig.~\ref{fig2}
while they completely diminish outside the boundary (shell region)
\cite{Note1}. This tendency becomes more pronounced with increasing the
attractive interaction and somewhat saturated in the strong limit. This
suggests that the pair density and the superfluid order (phase rigidity)
locally develop inside the core. This does not contradict a consensus
that the superfluid correlation grows as the most dominant one in 1-D
attractive Hubbard model \cite{1DHubbard} although any correlations
algebraically decay in 1-D systems. Thus, we call the core region
``superfluid core'' in the following. However, we note that there is a
subtle issue whether the whole core really exhibits superfluidity in
non-zero temperature due to its dimensionality and finiteness or
not. The discussion originates from a problem whether the phase rigidity
grows and survives from an edge of core to another edge against
fluctuations. This is beyond information obtained by the present
analysis of the exact diagonalization and DMRG. Thus, we do not discuss
the stability of the superfluidity but concentrate only on atom density
profiles in the ground state in this paper. We believe that the future
experiments and more advanced theoretical and numerical methods resolve
the problem (e.g., see Ref.~\cite{DMRGFF} for the static non-local
correlation of the pair function inside the core phase).  

\begin{figure}
\includegraphics[scale=0.33]{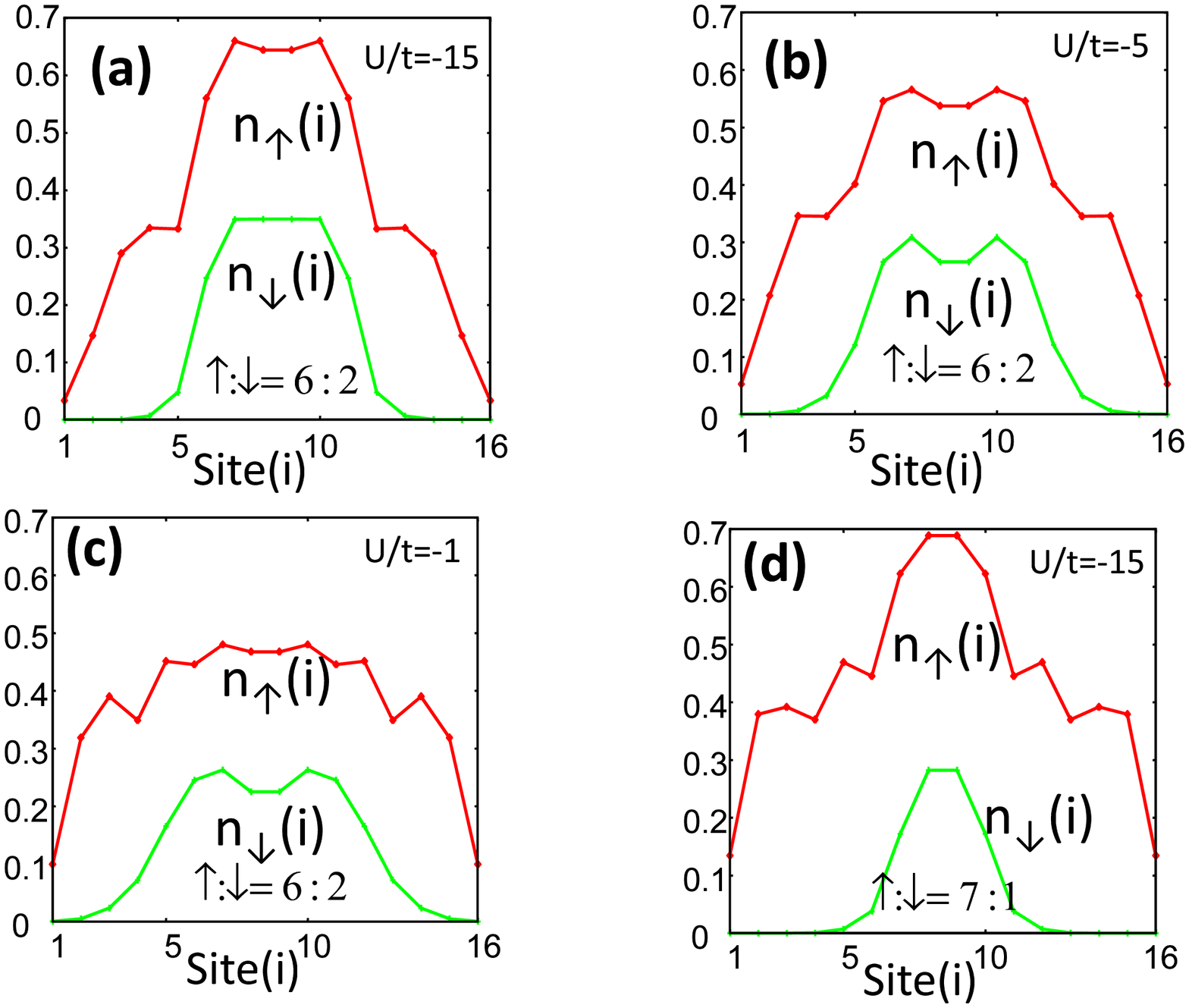}
\caption{\label{fig4} The exact diagonalization results of
 $n_{\uparrow}(i)$ and $n_{\downarrow} (i)$ profiles for (a) $U/t =
 -15$, (b) $-5$, and (c)$-1$. The numerical condition is the same as the
 case of Fig.2(b), in which the imbalance ratio is $3:1$. (d)
 $n_{\uparrow}(i)$ and $n_{\downarrow} (i)$ profiles for $U/t =-15$ in
 the case of the imbalance ratio $7:1$.} 
\end{figure}

Let us turn back to atom density profile in the presence of lattice,
again. Fig.~\ref{fig3} (a) reveals that un-paired component stays even
in the core region, where the $n_{\rm B}(i)$ and $\Delta(i)$ grow. As a
result, the density difference $\delta n(i)\equiv
n_\uparrow(i)-n_\downarrow(i)$ does not vanish in the core region as
seen in Fig.~\ref{fig3} (c). Generally, we find that
$n_\uparrow(i):n_\downarrow(i) \simeq 2:1$ in the core region except for
a very weak coupling regime (compare Fig.~\ref{fig2}(b) with
Fig.~\ref{fig3}(c)). This feature is clearly seen in Fig.4(a-c), which
display slices of $n_{\uparrow}(i)$ and $n_\downarrow (i)$ profiles for
$U/t = -15$, $-5$, and $-1$, respectively. Inside this range, it is
found that the ratio is approximately $2:1$. Also, the ratio is seen for
other imbalance cases (see e.g., Fig.~\ref{fig4} (d)), although there is
a small deviation from $2:1$. This is quite different from the
experimental result in the absence of the lattice, where one obtains
$n_\uparrow(i):n_\downarrow(i) \simeq 1:1$, or $\delta n(i)\simeq 0$ in
the core region \cite{Shin1,Part2}. 

\subsection{DMRG Studies}

\begin{figure}
\includegraphics[scale=0.42]{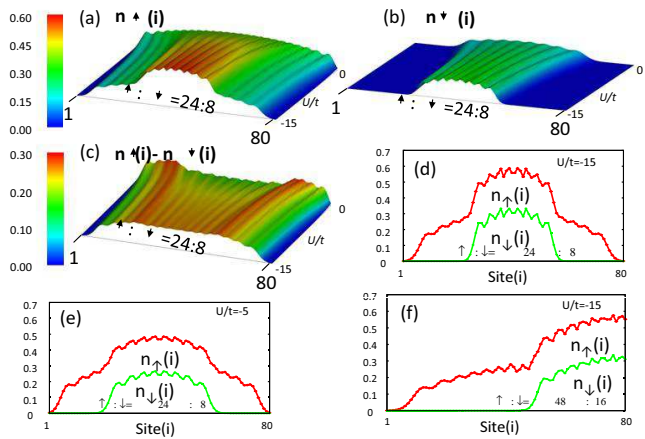}
\caption{\label{fig5} The DMRG results of $U/t$ dependences ($-15 \leq
U/t \leq 0$) for density profiles for (a) major species
 $n_{\uparrow}(i)$, (b) minor ones $n_{\downarrow}(i)$, and (c) the
 subtraction from major to minor one $n_{\uparrow} (i) - n_{\downarrow}
 (i)$. In these panels, $N=80$, and $N_{\rm F} = 32$ $(24 \uparrow, 8
 \downarrow)$. The panel (d) and (e) are two slices for $U/t =-15$ and
 $-5$ for $n_{\uparrow} (i)$ and $n_{\downarrow}(i)$, and the panel (f)
 is a half left-side profile of the local density of $n_{\uparrow} (i)$
 and $n_{\downarrow}(i)$ for $N=160$, $N_{\rm F} = 62$ $(\uparrow 48,
 \downarrow 16)$, and $U/t =-10$. In all cases, $V/t=1$.} 
\end{figure}

In order to confirm the generality of the phase separation and the core
phase characterized by the approximate ratio ``$2:1$,'' we adopt the
DMRG method and study much larger systems ($N=80$ and $160$). 
Figure~\ref{fig5} (a) and (b) show $U/t$ dependences ($-15 \leq U/t \leq 
0$) of density profiles for major and minor species, respectively, and
Fig.~\ref{fig5} (c) is their subtraction. In their results, the lattice
size $N=80$, $N_F=32$, and the prepaired ratio of different atom species
is $3:1$. It is found from Fig.~\ref{fig5} (a-b) that the atomic density
ratio in the core region is about ``$2:1$'' except for the weak coupling
region. The Fig.~\ref{fig5} (d) and (e) are two slices ($U/t=-15$ and
$-5$) of profiles of $n_\uparrow$ and $n_\downarrow$, which clearly
exhibit the approximate ratio ``$2:1$''. In addition, Fig.~\ref{fig5}
(f), which shows DMRG result (a half profile) of $n_\uparrow$ and
$n_\downarrow$ for $N=160$ and $N_{\rm F} = 64$, is also a clear
demonstration of the approximate ratio ``$2:1$.'' In these cases, it is
noted that the trap potential and the resulting confinement is
relatively very weak compared to the exact diagonalization results. From
the common results by both numerical methods, the approximate ratio
$n_{\uparrow} :n_{\downarrow} \simeq 2:1$ is widely expected to be
observed in the strong coupling regime in the presence of the optical
lattice with harmonic traps.  
 
\subsection{Effective Model and Discussion}

Both the exact diagonalization and the DMRG reveal that the excess
un-paired atoms remain even inside the core region in contrast to the
experimental results without the optical lattice. Especially, the ratio
of majority to minority species is approximately ``2:1'' inside the core
in the strong coupling regime. This result is considered to clearly
originate from strong correlation effects due to the presence of
lattice. Here, we derive an effective model in order to explain the
findings. In the strong-coupling limit $|U|\to\infty$, one finds that
the present lattice Fermi gas is effectively described by a gas mixture
of unpaired free Fermi-atoms and dimer atoms \cite{Machida1,Randeria}. 
The effective Hamiltonian has the form, (dropping the potential term) 
\begin{align}
H_{\rm Hubbard} 
&\to 2t^2 /U \sum_i ( n_{i+1} n_i -   B^\dag_{i+1} B_i ) 
\nonumber \\
& \quad {} - t \sum_i ( a^\dag_{i+1,\uparrow} a_{i,\uparrow} + {\rm
 H.c.}) \nonumber \\
& t \sum_i (a^\dag_{i+1,\uparrow}B_{i+1} B^\dag_{i} a_{i,\uparrow} +
 {\rm H.C.} ) \, . \label{eq.3} 
\end{align}
Here, $B_i^\dag$ is a creation operator of a bound pair at the $i$-th
site. In this effective Hamiltonian, the first term gives the
nearest-neighbor repulsive interaction $U_{\rm eff} \equiv 2t^2/U$
between Cooper-pair bosons and the energy gain of the molecular hopping
$-t_{\rm eff} \equiv -2t^2/U$ between the nearest-neighbor sites. The
second term in Eq.~(\ref{eq.3}) stands for the kinetic energy of excess
$\uparrow$-spin atoms, and the third term describes an exchange energy
between a Cooper pair and an excess atom when they sit on the nearest
neighbor position each other. The third term is also derived by
considering the hopping of the minor atom from an on-site Cooper pair to
the nearest neighbor site, in which an excess major atom solely
stays. This effective model is valid in $|U|\to\infty$, since there
exists conversion process between the bosons and the excess fermions for
finite negative $U/t$. However, it is numerically found that the limit
description ($|U| \to \infty $) becomes sufficiently good more than
$|U/t|\sim 10$, since the un-paired fermions almost disappear in the
range. See Ref.~\cite{Machida1} for $U/t$ dependence of the total number
of un-paired fermions $( \equiv \sum_i n_F (i) )$ in the balanced cases.   

In the effective model considering only the first term of
Eq.~(\ref{eq.3}), i.e., in the balanced cases, it is well-known that the
pair density correlation function shows a large magnitude between every
other sites. The reason is that the Cooper pair tends to occupy every
other site to avoid energy loss by the nearest-neighbor repulsion
between Cooper-pair bosons (see the first term of Eq.~(\ref{eq.3})). As
a result, such a correlation effect causes a static spatial modulation
of the pair density in the presence of the trap, which breaks the
translational symmetry. In other words, a degeneracy between two
dominant states having the site density correlation of pairs as ``$-
n_{\uparrow \!\downarrow} - 0 - n_{\uparrow \! \downarrow} - 0 -$'' and
``$- 0 - n_{\uparrow \! \downarrow} - 0 - n_{\uparrow \! \downarrow} -
$'', where $n_{\uparrow \! \downarrow}$ is the site density of the
Cooper pair ($n_{\uparrow\downarrow}<1$), is broken by the trap
potential, and the static density modulation as represented by ``$-
n'_{\uparrow \! \downarrow} - n''_{\uparrow \! \downarrow} -
n'_{\uparrow \! \downarrow} - n''_{\uparrow \! \downarrow}$''
appears. This behavior in the balanced case was also confirmed by
different authors \cite{Machida1,Gao,Molina}. 

On the other hand, in the imbalanced systems, both the first and third
terms of Eq.~(\ref{eq.3}) explain the present numerical results of the
density profiles. The first term prefers two degenerate density
correlations described above, and excess atoms can then find their
positions in the nearest neighbor sites of the on-site
pair. Furthermore, the third term stabilizes the two density-correlation
profiles, in which excess majority atoms slip in like ``$- n_{\uparrow
\!\downarrow} - n_{\uparrow} - n_{\uparrow \! \downarrow} - n_{\uparrow}
- $'' and ``$- n_{\uparrow} - n_{\uparrow \!\downarrow} - n_{\uparrow} -
n_{\uparrow \! \downarrow} - $'', since the energy $2t$ is gained by an
anti-bonding exchange between an on-site pair and its neighbor excess
atom. As a result, one finds that the density difference in the core
region is simply given by $n_\uparrow:n_\downarrow \simeq 2:1$ in the
large $|U|$ range. We emphasize that the presence of the lattice is
responsible for the approximate ratio $2:1$ while it is absent in a
Fermi gas with no lattice potential.

\subsection{Strong Trap Effect}

\begin{figure}
\includegraphics[scale=0.4]{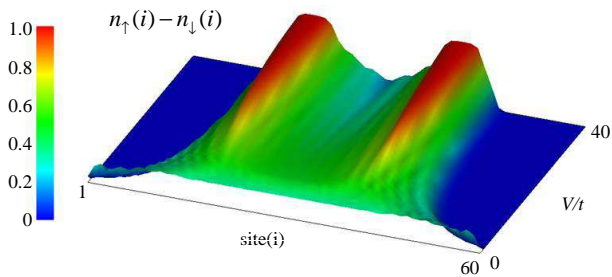}
\caption{\label{fig6} The DMRG results of $V/t$ dependence of the
particle density profile for $N=60$, $N_{\rm F} = 40$ $(30\uparrow,
10\downarrow)$, and $U/t = -10$.}
\end{figure}

In this subsection, we study a confinement force dependence on the atom
density profile. Figure \ref{fig6} shows $V/t$ dependence on the density
profile for the imbalance ratio $N_\uparrow: N_\downarrow =
3:1$. The focus of interest is a case in which the strength of the trap
potential is so strong that the confinement force overwhelms the
correlation effects due to the presence of lattice, i.e., the repulsive
interaction between Cooper pairs and the exchange energy gain between a
Cooper pair and an excess atom as explained above. Such an extreme
situation is observable above $ V/t\sim 35 $ as seen in Fig.~\ref{fig6}. 
One finds above $ V/t\sim 35 $ that un-paired atoms are excluded from a
central part of the core region. Then, the density difference from major
to minor species becomes $\delta n(i)\simeq 0$, namely, the ratio
$n_\uparrow:n_\downarrow \simeq 1:1$. This feature is equivalent with
the recent experiments without lattices \cite{Shin1,Part2}. However,
such a situation may be non-realistic, since it requires very large $V$
compared to the non-lattice one. Moreover, the numerical simulation
result (Fig.~\ref{fig6}) reveals that only a central point of the core
region shows $n_\uparrow:n_\downarrow=1:1$. This is also not
experimentally observable in non-lattice cases. It is found even in
strongly confined cases that the lattice presence is non-negligible.

\section{Numerical Results for 2-D Imbalanced Gases\label{NR2D}}

In this section, we study 2-D imbalanced gases in the presence of the
optical lattice. The lattice is a square one, in which only the nearest
neighbor hopping is allowed. In this paper, we employ only the exact
diagonalization. Thus, the total number of lattice sites $N$ is limited
to 25, and $N_F =12$, where the imbalance ratio $n_\uparrow:n_\downarrow
= 3:1$ $(9 \uparrow, 3 \downarrow)$.
    
\begin{figure}
\includegraphics[scale=0.33]{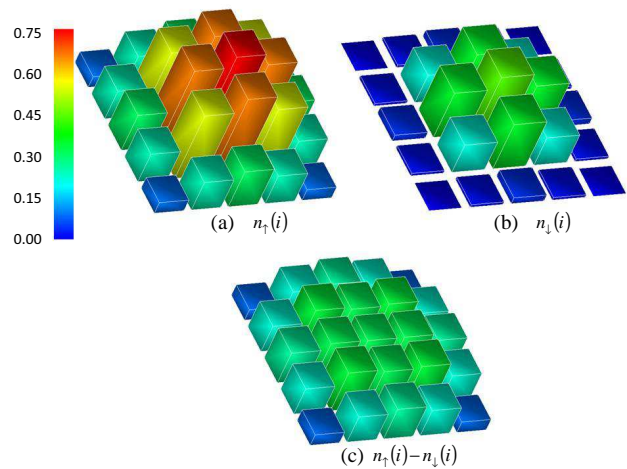}
\caption{\label{fig7} The exact diagonalization results of the particle
 density profiles of (a) $n_{\uparrow}(i)$, (b) $n_{\downarrow}(i)$ and  
(c) $n_{\uparrow} (i) -n_{\downarrow} (i)$ for 2-D attractive imbalanced 
Hubbard model with a 2-D harmonic trap potential. In this case, $N =
 25$, $N_{\rm F} = 12$ $(9 \uparrow, 3 \downarrow )$, $U/t = -15$, and
 $V/t = 1$.}
\end{figure}

Figure \ref{fig7} shows a typical result of the calculated density
profiles for $n_\uparrow(i)$, $n_\downarrow(i)$, and their subtraction
$n_\uparrow(i) - n_\downarrow(i) $ in a two-dimensional lattice Fermi
gas, in which $5 \times 5$ $(=25)$ square lattice is used, and $U/t$ and 
$V/t$ are taken $-15$ and $1$, respectively. Although the system size is
small, we find $n_\uparrow:n_\downarrow \simeq 2:1$ inside the core,
again. This result indicates that the penetration of un-paired excess
atoms into the core region is a general feature in the presence of a
lattice, irrespective of dimensionality of the system. 

\section{Finite Temperature Effects\label{TempE}}

Finally, let us mention finite temperature effects on the approximate
ratio 2:1 in the core phase. We find that there are two crucial
temperatures in discussing the temperature effects on the present
results. The first temperature, which is much higher than the second
temperature, is given by an energy scale in which the Hubbard model
becomes sufficiently valid, i.e.,  the single-band Hubbard model is a
good description. The experiments have already succeeded in reaching a
range below the first temperature. The second temperature is
characterized by a lower energy range that the correlation effects have
an important role, i.e., the present features become effective. Thus,
the problem is whether one can cool down below the second temperature or
not. Let us estimate the second temperature. In the half-filling of 1-D
lattice, $0.5t \sim E_{\rm F}$, where $E_{\rm F}$ is the Fermi energy
and $t$ is the the hopping matrix element in the Hubbard model, while
the energy scale for the main correlation effect is $t^2 / U$ as shown
in Eq.~(\ref{eq.3}) in this manuscript. Thus, if one assumes $U/t \sim
10$, then the energy scale for the correlation $\sim 0.1t$. This means
that the correlation effect becomes effective when the temperature is
lower than $0.2 E_{\rm F}$. The temperature range $T < 0.2 E_{\rm F}$
was achieved in the absence of the optical lattice.

\section{Summary}

We investigated ground state properties of two-component trapped Fermi 
gases with population imbalances loaded on optical lattices. Using the
exact diagonalization method and partly DMRG, we calculated the density
profiles for 1-D and 2-D Hubbard model, as a function of the attractive
interaction for various population imbalances. We showed that the
population imbalance remains in the core phase, where the ratio of major
to minor species, $n_\uparrow:n_\downarrow \simeq 2:1$. This result is
observed in the wide parameter range except for the weak attractive
interaction or the extremely strong trap one irrespective of the
dimensionality. This is different from the case in the absence of the
lattice, where $n_\uparrow :n_\downarrow=1:1$ is observable in the core
region. The character in the core phase on the optical lattice can be
explained by the correlation effects on the lattice through the
effective model. 

{\it Note added} --- Recently, we became aware of
Refs.~\cite{Torma,DMRGFF}, which analyze the same model and give
consistent results with ours.

\begin{acknowledgments}
 M.M. would like to thank T.~Koyama, M.~Kato, T.~Ishida, H.~Ebisawa, and 
 N.~Hayashi for helpful discussions about superconductivity. He also
 thanks Y.~Morita for his support, and T.~Imamura, T.~Kano and staff
 members in the Earth Simulator center for their supports in large-scale
 calculations. Two of authors (M.M. and S.Y.) acknowledge M.~Kohno,
 T.~Hotta, and H.~Ohnishi for illuminating discussion about the DMRG
 techniques. The work was partially supported by Grant-in-Aid for
 Scientific Research on Priority Area ``Physics of new quantum phases in
 superclean materials'' (Grant No.~18043022 and 18043005) from the
 Ministry of Education, Culture, Sports, Science and Technology of
 Japan. In addition, this work was supported by Grant-in-Aid for
 Scientific Research from MEXT, Japan (Grant No.~16740187, 17540368,
 18500033, 18540387, and 17540368), and performed under a support by
 JSPS Core-to-Core Program-Strategic Research Networks ``Nanoscience and
 Engineering in Superconductivity (NES)''. 
\end{acknowledgments}


\begin{references}
\bibitem{NoteD} Here, we note that $n_\uparrow :n_\downarrow = 2:1$ does
 not mean the double and the single occupations of a site by atoms,
 respectively. An atom is generally delocalized over several
 sites. Therefore, $n_\sigma (i)$ can become less than 1 if the filling
 is below the half-filling without the trap potential. The ratio 2:1 in
 the core region as shown in Fig.~\ref{fig1} (b) indicates the ratio of
 the observable atomic densities on a single site.
\end{references}
\end{document}